\documentclass[twocolumn,secnumarabic,amssymb, nobibnotes, aps, prd,showpacs]{revtex4}
\usepackage{amssymb}
\usepackage{amsmath}
\usepackage{array}
\usepackage{graphicx}
\usepackage{dcolumn}
\usepackage{bm}
\usepackage{multirow}
\usepackage{float}
\newcommand{\dtabsize}{\tiny}
\newcommand{\clee}{C_{\ell}^{EE}}

\begin{document}


\title{Principal component analysis of the reionization history from Planck 2015 data}

\author{Wei-Ming Dai}
\email{daiwming@itp.ac.cn}
\affiliation{State Key Laboratory of Theoretical Physics, Institute of
Theoretical Physics, Chinese Academy of Sciences, P.O. Box 2735,
Beijing 100190, China}

\author{Zong-Kuan Guo}
\email{guozk@itp.ac.cn}
\affiliation{State Key Laboratory of Theoretical Physics, Institute of
Theoretical Physics, Chinese Academy of Sciences, P.O. Box 2735,
Beijing 100190, China}

\author{Rong-Gen Cai}
\email{cairg@itp.ac.cn}
\affiliation{State Key Laboratory of Theoretical Physics, Institute of
Theoretical Physics, Chinese Academy of Sciences, P.O. Box 2735,
Beijing 100190, China}


\begin{abstract}
The simple assumption of an instantaneous reionization of the Universe may bias estimates of cosmological parameters.
In this paper a model-independent principal component method for the reionization history is applied to give constraints on the cosmological parameters from recent Planck 2015 data.
We find that the Universe are not completely reionized at redshifts $z \ge 8.5$ at 95\% CL.
Both the reionization optical depth and the matter fluctuation amplitude are higher than but consistent with those obtained in the standard instantaneous reionization scheme.
The high estimated value of the matter fluctuation amplitude strengthens the tension between Planck CMB observations and
some astrophysical data, such as cluster counts and weak lensing.
The tension can significantly be relieved if the neutrino masses are allowed to vary.
Thanks to a high scalar spectral index, the low-scale spontaneously broken SUSY inflationary model can fit the data well,
which is marginally disfavored at 95\% CL in the Planck analysis.
\end{abstract}


\maketitle


\section{Introduction}

Accurate measurements of the reionization history of the Universe plays an important role
in understanding the early stages of structure and star formation.
Fortunately, the evolution of the inter-galactic Lyman-alpha opacity measured in the spectra of quasars
can provide valuable information on the reionization history~\cite{Gunn:1965hd}.
The recent measurements suggest that the reionization of the inter-galactic medium
was largely complete at redshift $z\approx 6$~\cite{Fan:2005es}.
The rapid decline in the space density of Lyman-alpha emitting galaxies in the region $z=6-8$
also implies a low-redshift reionization process~\cite{Choudhury:2014uba}.
However, the detailed evolution of the electron ionization fraction with redshift is currently unknown
due to the difficulties in probing the reionization state of the Universe at higher redshifts through the Lyman-alpha opacity.
The most promising observation in the near future is detection of the 21 cm transition of neutral hydrogen.
The 21 cm emission is a direct probe of the large-scale distribution of neutral hydrogen during the epoch of reionization,
and is, therefore, a complementary probe to the Lyman-alpha emission at lower redshifts.
According to current understanding of reionization sources,
high energy photons may arise from the first star-forming galaxies and quasars.
The uniqueness of these objects as the sources of photons to reionize the Universe is more challenging to establish.

High precision measurements of the cosmic microwave background (CMB) anisotropies provide new insights into the reionization history.
There are two main effects of reionization on the CMB angular power spectra.
The first effect produces a suppression of the acoustic peaks in the CMB angular power spectra
through the optical depth integrated over the whole reionization history.
Earlier reionization leads to the larger suppression of the acoustic peaks.
The second effect changes the shape of the large-scale polarization angular power spectra.
This gives a characteristic bump in the spectra on scales larger than the horizon size at reionization.
The position of the bump is proportional to the square root of the redshift at which the reionization occurs,
while the amplitude is proportional to the optical depth~\cite{Zaldarriaga:1996ke,Hu:1996mn}.
Therefore, measurements of the large-scale polarization angular power spectra can
be used to constrain the reionization history~\cite{Kaplinghat:2002vt,Hu:2003gh}.

Current constraint on the reionization optical depth from the Planck temperature data in combination with the low-multipole polarization data
is $\tau=0.078\pm 0.019$ in the base $\Lambda$CDM model~\cite{Ade:2015xua},
which is smaller than $\tau=0.089\pm 0.014$ estimated by the WMAP 9-year data~\cite{Bennett:2012zja},
but consistent with that from WMAP polarization measurements cleaned for dust emission
using 353 GHz polarization maps from the high frequency instrument~\cite{Ade:2013zuv}.
These results are based on the assumption of an instantaneous reionization.
If such a simple assumption is not true, the estimated values of cosmological parameters are biased.
Since there is currently no convincing model of the reionization history,
it is important and necessary to constrain it in a relatively model-independent way.

Hu and Holder proposed a principal component analysis (PCA) of the reionization history to
quantity the information contained in the large-scale $E$-mode polarization~\cite{Hu:2003gh}.
The principal components are the eigenfunctions of the Fisher matrix
that describes the dependence of the polarization angular power spectra on the reionization history.
In practice, no more than the first five principal components are needed to
parameterize the reionization history~\cite{Mortonson:2007hq}.
This approach has been applied to WMAP data and simulated future
data~\cite{Mortonson:2007tb,Colombo:2008jr,Mortonson:2008rx,Mortonson:2009qv,Archidiacono:2010wp,Pandolfi:2011kz}.
It is recently claimed  that the Harrison–-Zel'dovich primordial spectrum is consistent with the WMAP 7-year data
in such a general reionization scenario~\cite{Pandolfi:2010dz,Pandolfi:2010mv}.
In this paper, we perform the principal component analysis of the reionization history from recently released Planck 2015 data
and study the impact of the reionization history on the estimates of the cosmological parameters.
We also investigate the bounds on the neutrino masses and primordial tensor perturbations in a general reionization scheme.

The paper is organized as follows.
In Section~\ref{sec:model} we describe the principal component method for parameterizing the reionization history
that we will adapt in our analysis.
The results are presented in Section~\ref{sec:results}.
The last section is devoted to our conclusions.

\section{Model}\label{sec:model}

Since there is no convincing physical model of the reionization history,
the instantaneous reionization history is usually assumed when investigating cosmological constraints from CMB measurements.
Such a one-parameter phenomenological model may bias the estimates of the optical depth and other cosmological parameters.
In general, the hydrogen ionization fraction $x_e(z)$ is a function of redshift.
Binning the hydrogen ionization fraction in redshift bins is a good idea to parameterize the reionization history.
However, the degeneracy between the binned ionization fraction leads to weak constraints~\cite{Lewis:2006ym}.
The principal component analysis provides a solution to this problem~\cite{Hu:2003gh}.

The PCA approach is often adopted to reduce the number of variables while containing most of information of the system.
It converts a set of correlated variables into a set of linear uncorrelated variables by an orthogonal transformation.
The principal components are the eigenvectors of the diagonalized matrix, which are picked out according to the corresponding eigenvalues.
Intuitively, the larger the eigenvalue is, the more information the corresponding eigenvector gives.

\begin{figure}
\centering
\includegraphics[width=3in,height=2.5in]{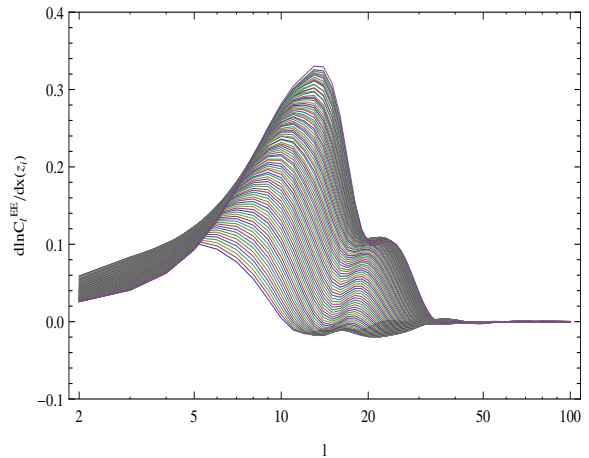}
\caption{Fractional power spectrum response to a delta function perturbation of unit amplitude at $6.25<z_i<29.75$.
Here we have chosen $\Delta z=0.25$.}
\label{fig:DiffCee}
\end{figure}

Following~\cite{Mortonson:2007hq},we  parameterize the reionization history.
Consider a binned ionization fraction $x_e(z_i)$, $i\in \{1,2,\ldots,N_z\}$, 
with redshift bins of width $\Delta z=0.25$ spanning $z_{\rm min}\leq z \leq z_{\rm max}$.
The bin width of $\Delta z=0.25$ is sufficiently small 
so that the final results we obtain are independent of the redshift binning~\cite{Mortonson:2007hq}.
Here we take $z_{\rm min}=6$ and $z_{\rm max}=30$.
With the definition $z_1=z_{\rm min}+\Delta z$ and $z_{N_z}=z_{\rm max}-\Delta z$ so that $N_z+1=(z_{\rm max}-z_{\rm min})/\Delta z$.
For $z \geq z_{\rm max}$, the ionization fraction is set to the residual fraction of recombination,
while $x_e = 1.0$ for $3 \leq z\leq z_{\rm min}$ consistent with astrophysical observations~\cite{Fan:2005es}
and $x_e=1.08$ for $z < 3$ when the Helium reionization is taken into account.
The principal components of $x_e(z_i)$ are the eigenfunctions of the following Fisher matrix $F_{ij}$,
\begin{equation}
F_{ij}=\sum_{\ell=2}^{\ell_{\rm max}}\left(\ell+\frac{1}{2}\right)
       \frac{\partial \ln \clee}{\partial x_e(z_i)}
       \frac{\partial \ln \clee}{\partial x_e(z_j)} \,,
\label{eq:fisher1}
\end{equation}
which describes the dependence of the polarization spectrum $\clee$ on the ionization fraction $x_e(z_i)$.
To calculate the Fisher matrix, we should set a fiducial reionization history $x_e^{\rm fid}(z_i)$ as well as other cosmological parameters. 
The fiducial model parameters are chosen as the estimated values of cosmological parameters from the Planck 2015 data
including the temperature and polarization power spectra for the base $\Lambda$CDM model~\cite{Ade:2015xua},
listed in Table~\ref{tab:fiducial}.
We adopt $x_e^{\rm fid}(z_i)=0.1$ in the range of redshifts $6\le z \le 30$ to get the same optical depth as the instantanous reionization.
Actually our results are not sensitive to the fiducial model.
Since the effect of $x_e(z_i)$ on $\clee$ of can be neglected on high $\ell$, as shown in Fig.~\ref{fig:DiffCee}, we can cut off $\ell$ safely to $\ell_{\rm max}=100$ in the Fisher matrix.
The perturbation to the fiducial $x_e^{\rm fid}$ is a delta function correlated with the nearby region~\cite{Hu:2003gh}.

\begin{table}[!htbp]
\begin{center}
\caption{Fiducial model consistent with results from the 2015 Planck temperature and polarization spectra
for the standard $\Lambda$CDM model.}
\begin{tabular}{cccccc}
\hline\hline
$\Omega_b{h}^2$ &$\Omega_c{h}^2$ &h & $\tau$ & $A_se^{-2\tau}$ & $n_s$ \\ \hline
0.02225 &0.1198 &0.6727 & 0.079 & $1.882\times10^{-9}$ &0.9645 \\ \hline
\end{tabular}
\label{tab:fiducial}
\end{center}
\end{table}

\begin{figure}[H]
\begin{center}
\includegraphics[width=3in]{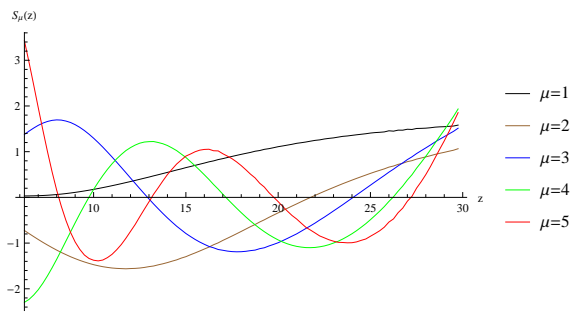}
\caption{The five lowest-variance principal components of $x_e(z)$ over the redshift range of $6.25<{\rm z}<29.75$.
The variances increase with the index $\mu$.}
\label{fig:pca_bases}
\end{center}
\end{figure}

The Fisher matrix $F_{ij}$ is decomposed as
\begin{equation}
F_{ij}=(N_z+1)^{-2}\sum_{\mu=1}^{N_z}S_\mu(z_i)\sigma_\mu^{-2}S_{\mu}(z_j) \,,
\label{eq:fisher2}
\end{equation}
where $\sigma_\mu^2$ are the inverse eigenvalues and
$S_\mu(z)$ are the eigenfunctions that satisfy the orthogonality and completeness relations
\begin{eqnarray}
\int_{z_{\rm min}}^{z_{\rm max}} dz S_\mu(z)S_\nu(z)&=&(z_{\rm max}-z_{\rm min})\delta_{\mu\nu} \,,
\label{eq:orthog} \\
\sum_{\mu=1}^{N_z} S_\mu(z_i)S_\mu(z_j)&=&(N_z+1)\delta_{ij} \,.
\label{eq:compl}
\end{eqnarray}
So these eigenfunctions form a set of orthogonal and complete bases.
An arbitrary reionization history can be decomposed into a sum of the eigenmodes, 
\begin{equation}
x_e(z)=x_e^{\rm fid}(z)+\sum_\mu m_\mu S_\mu(z) \,,
\label{eq:pcs}
\end{equation}
where $m_\mu$ are the amplitudes of the principal components for a particular reionization history.
We order the eigenmodes so that the smallest $\sigma_\mu^2$ has the lowest index number $\mu$, starting at $\mu=1$.
Since the effect of each eigenmode on $\clee$ becomes smaller as $\mu$ increases,
we shall pick out the first few eigenmodes.
It is enough to choose the first $3-5$ eigenmodes shown in Fig.~\ref{fig:pca_bases}
because of the existence of the cosmic variance
\begin{equation}
\frac{\Delta \clee}{\clee} = \sqrt{\frac{2}{2 \ell +1}} \,,
\label{eq:cosmicVar}
\end{equation}
so that the error in $\clee$ from high $\mu$ eigenmodes is smaller than the cosmic variace.

In our analysis, we parameterize the reionizaton history as~\eqref{eq:pcs} with the five amplitudes $m_\mu$ for $\mu =1,...,5$.
We consider a spatially-flat $\Lambda$CDM model described by a set of cosmological parameters
\begin{equation*}
\{\Omega_bh^2,\Omega_ch^2,\Theta_s,A_s,n_s\},
\end{equation*}
where $\Omega_bh^2$ and $\Omega_ch^2$ are the physical baryon and cold dark matter densities relative to the critical density,
$\Theta_s$ is the ratio of the sound horizon to the angular diameter distance at decoupling,
$A_s$ is the amplitude of the primordial power spectrum of scalar perturbations
and $n_s$ is the scalar spectral index.
Moreover, we consider two one-parameter extensions to the $\Lambda$CDM model,
$\Lambda$CDM+$\Sigma m_\mu$ and $\Lambda$CDM+$r$,
where $\Sigma m_\mu$ is the total mass of neutrinos and $r$ is the tensor-to-scalar ratio.
When the tensor contribution to the CMB angular power spectra is considered,
the inflationary consistency relation, $n_t=-r/8$, is assumed
where $n_t$ is the tensor spectral index.
For comparison with the Planck results, $r$ is defined at the pivot scale $k_*=0.002$ Mpc$^{-1}$ (denoted $r_{0.002}$).

We modified the Boltzmann CAMB code~\cite{Lewis:1999bs} to appropriately incorporate
the PCA reionization history and
used the publicly available CosmoMC package to explore the parameter space by means of Monte Carlo Markov chains technique~\cite{Lewis:2002ah}.

In our analysis we use recently released Planck likelihood code and data,
including the Planck low-$l$ likelihood at multipoles $2\le l \le 29$ and
high-$l$ Plik likelihood at multipoles $l\ge 30$ based on pseudo-$C_l$ estimators.
The former uses foreground-cleaned LFI 70 GHz polarization maps together with the temperature map obtained from the Planck 30 to 353 GHz
channels by the Commander component separation algorithm over 94\% of the sky.
The polarization part of this likelihood is denoted as ``lowP''.
The latter uses 100 GHz, 143 GHz, and 217 GHz half-mission cross-power spectra,
avoiding the Galactic plane as well as the brightest point sources and the regions where the CO emission is the strongest.
``Planck TT+lowP'' denotes the combination of the TT likelihood at $l \ge 30$ and a low-$l$ temperature-polarization likelihood.
``Planck TT,TE,EE+lowP'' denotes the combination of the likelihood at $l \ge 30$ using TT, TE,
and EE spectra and the low-$l$ temperature-polarization likelihood.

\section{Results}\label{sec:results}

Table~\ref{tab:parameters} summarizes the constraints on the reionization history and cosmological parameters
from the Planck TT+lowP and Planck TT,TE,EE+lowP data.
The results from the temperature and polarization power spectra are consistent with those from Planck TT+lowP,
but with increased precision.
Since temperature-to-polarization leakage corrections are not considered in the 2015 Planck TE and EE power spectra~\cite{Ade:2015xua},
in our analysis Planck TT+lowP is usually quoted.
Actually the reionization history is not sensitive to the TE and EE power spectra at $l \ge 30$.

\begin{table*}[!htmb]
\begin{center}
\caption{68\% constraints on the PCA reionization and other cosmological parameters from 2015 Planck CMB angular power spectra.
For the total mass of neutrinos and tensor-to-scalar ratio, the 95\% upper limits are given.}
\begin{tabular}{|>{\dtabsize}c| >{\dtabsize}c>{\dtabsize}c>{\dtabsize}c>{\dtabsize}c|}
\hline
\hline
\multirow{2}{*}{Parameters}    & $\Lambda$CDM   & $\Lambda$CDM &$\Lambda$CDM$+r$ &$\Lambda$CDM$+\Sigma{m_\nu}$  \\ \cline{2-5}
                               & Planck TT+lowP & Planck TT,TE,EE+lowP & Planck TT+lowP  & Planck TT+lowP     \\
\hline
$\Omega_{\mathrm{b}}h^2$      & $0.02223\pm0.00023$  & $0.02224\pm0.00016$  & $0.02225\pm0.00023$   & $0.02210\pm0.00030$  \\
$\Omega_{\mathrm{c}} h^2$     & 0.1192$\pm$0.0021    & $0.1196\pm0.0015$    & $0.1190\pm0.0021$     & $0.1201\pm0.0024$    \\
$100\theta_{\mathrm{MC}}$     & 1.04092$\pm$0.00047  & $1.04079\pm0.00032$  & $1.04095\pm0.00047$   & $1.04066\pm0.00056$  \\
$\tau$                        & 0.092$\pm$0.017      & $0.092\pm0.015$      & $0.092\pm0.017$       & $0.095\pm0.017$      \\
$n_\mathrm{s}$                & 0.9689$\pm$0.0064    & $0.9663\pm0.0049$    & $0.9698\pm0.0064$     & $0.9662\pm0.0075$    \\
$\ln(10^{10} A_\mathrm{s})$   & 3.115$\pm$0.033      & $3.117\pm0.030$      & $3.115\pm0.032$       & $3.121\pm0.034$      \\
$H_0$ (km s$^{-1}$ Mpc$^{-1}$)& 67.54$\pm$0.96       & $67.34\pm0.65$       & $67.64\pm0.96$        & $64.85\pm3.34$       \\
$\sigma_8$                    & $0.838\pm0.014$      & $0.840\pm0.012$      & $0.839\pm0.014$       & $0.788\pm0.063$      \\
$\Omega_m$                    & $0.3116\pm0.0132$    & $0.3144\pm0.0091$    & $0.3104\pm0.0131$     & $0.3507\pm0.0525$    \\
$r_{0.002}$                   & $-$                  & $-$                  & $<0.091$              & $-$                  \\
$\Sigma{m_\nu}$ (eV)          & $-$                  & $-$                  & $-$                   & $<1.03$              \\
$m_1$           & $0.0569^{+0.0625}_{-0.0534}$  & $0.0485^{+0.0590}_{-0.0510}$  & $0.0528^{+0.0608}_{-0.0523}$  & $0.0663^{+0.0712}_{-0.0563}$   \\
$m_2$           & $-0.0457^{+0.1193}_{-0.1091}$ & $-0.0689^{+0.1073}_{-0.1065}$ & $-0.0543^{+0.1198}_{-0.1076}$ & $-0.0536^{+0.1300}_{-0.1096}$  \\
$m_3$           & $0.0498^{+0.1450}_{-0.1440}$  & $0.0409^{+0.1338}_{-0.1489}$  & $0.0586^{+0.1381}_{-0.1340}$  & $0.0642^{+0.1465}_{-0.1613}$   \\
$m_4$           & $-0.010^{+0.1503}_{-0.1500}$  & $-0.0190^{+0.1499}_{-0.1398}$ & $-0.0127^{+0.1435}_{-0.1427}$ & $-0.0102^{+0.1549}_{-0.1547}$  \\
$m_5$           & $0.0524^{+0.1566}_{-0.1570}$  & $0.0406^{+0.1460}_{-0.1448}$  & $0.0519^{+0.1551}_{-0.1549}$  & $0.0560^{+0.1688}_{-0.1689}$   \\ \hline
\end{tabular}
\label{tab:parameters}
\end{center}
\end{table*}

\begin{figure}
\begin{center}
\includegraphics[width=2.5in,height=2in]{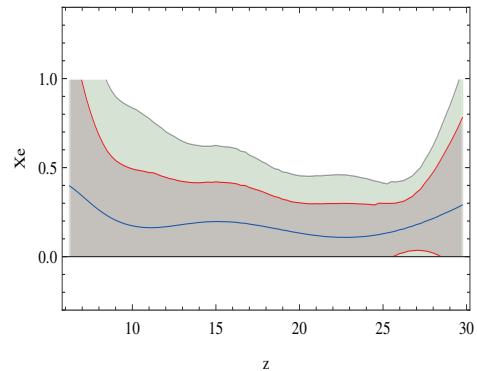}
\caption{Reconstructed reionization history with the PCA method from PlanckTT+lowP.
The blue curve is the mean value of the hydrogen ionization fraction $x_e(z)$ in the redshift range $6.25<z<29.75$,
with $68\%$ (dark color) and $95\%$ (light color) confidence regions.}
\label{fig:xe}
\end{center}
\end{figure}

\begin{figure*}
\begin{center}
\includegraphics[width=4.5in,height=4.5in]{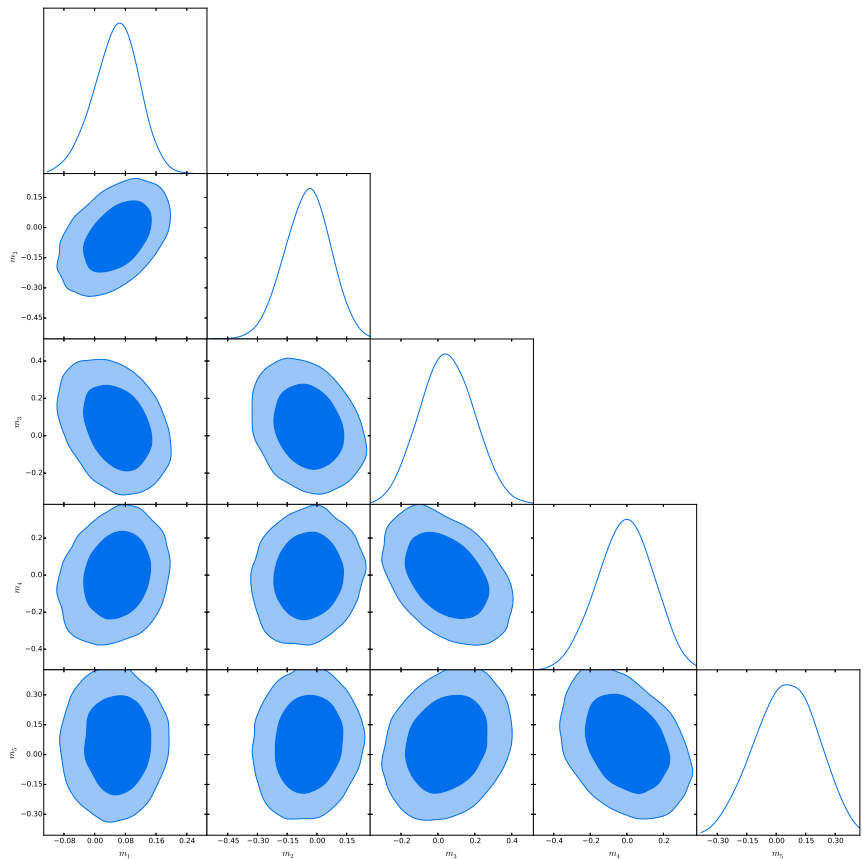}
\caption{Marginalized 2D contours (68\% and 95\% CL) and posterior distributions for the amplitudes of the principal components $m_{\mu}$, derived from Planck TT+lowP.}
\label{fig:tri}
\end{center}
\end{figure*}

The mean posterior values and 68\% confidence limits of $m_\mu$ are listed in Table~\ref{tab:parameters}.
Compared to WMAP data, Planck data place strong constraints on the amplitudes of the principal components.
The marginalized 2D contours and posterior probability distributions are plotted in Figure~\ref{fig:tri}.
As expected, the correlations between $\{m_\mu\}$ are weak and the probability distributions are Gaussian.
Moreover, we reconstruct the reionization history from Monte Carlo chains using the PCA approach in Figure~\ref{fig:xe}.
The blue curve is the mean value of the hydrogen ionization fraction $x_e(z)$ in the redshift range $6.25<z<29.75$,
with $68\%$ (dark color) and $95\%$ (light color) confidence regions.
As we can see the Universe is not completely reionized at $z \ge 8.5$ at 95\% CL.
This result is consistent with the estimated value of the reionization redshift $z_{\rm re}=8.8^{+1.7}_{-1.4}$
derived from Planck TT+lowP+lensing in the instantaneous reionization model.

\begin{figure*}
\includegraphics[width=2.5in,height=2.5in]{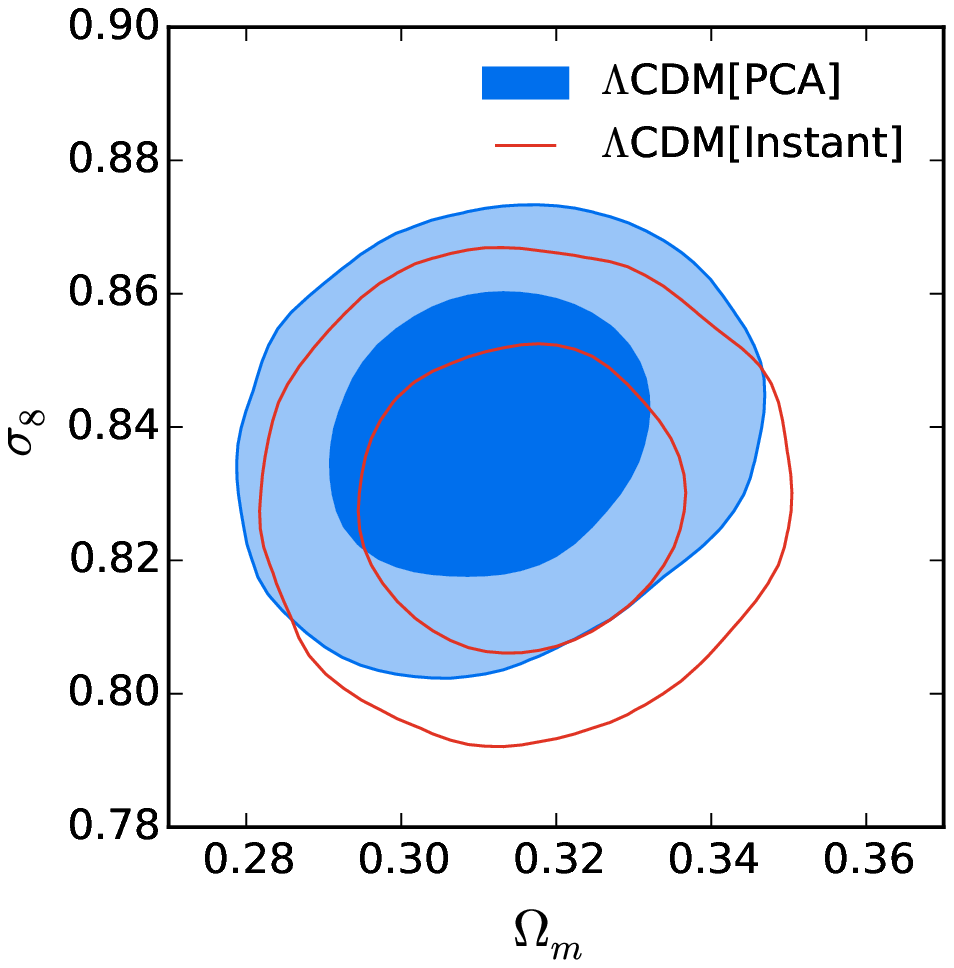}
\includegraphics[width=2.5in,height=2.5in]{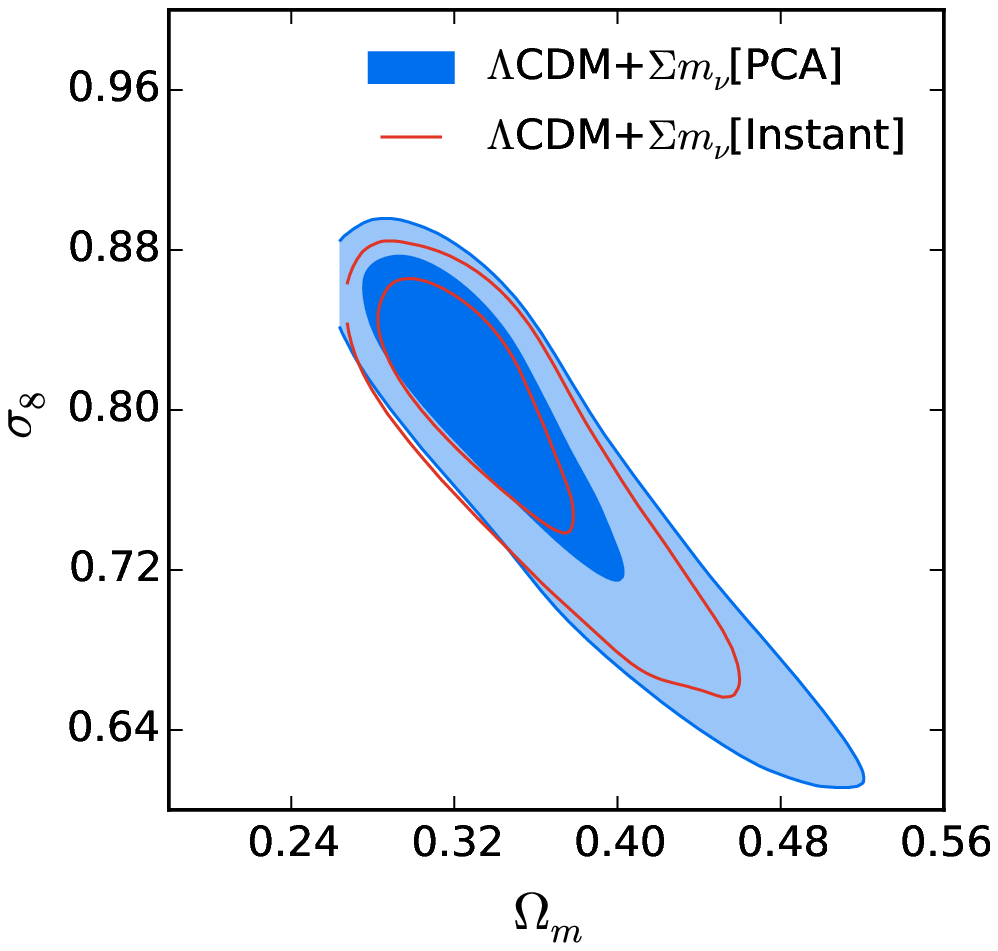}
\caption{Marginalized 2D contours in the $\sigma_8 - \Omega_m$ plane, with $68\%$ and $95\%$ CL,
for the $\Lambda$CDM model (left panel) and $\Lambda$CDM+$\Sigma m_\nu$ (right panel)
with the PCA (blue) and instantaneous (red) reionization history, derived from Planck TT+lowP.}
\label{fig:sigma8}
\end{figure*}

\begin{figure}
\includegraphics[width=2.5in,height=2.5in]{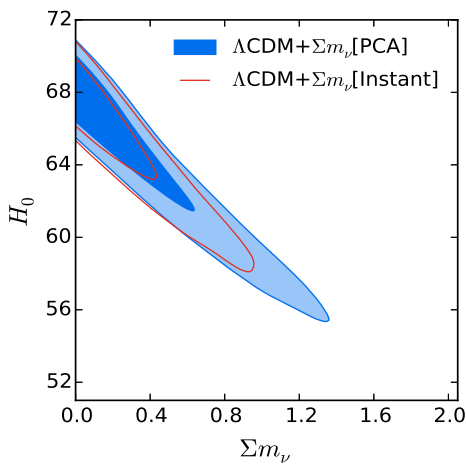}
\caption{Marginalized 2D contour in the $H_0 - \Sigma m_\nu$ plane, with 68\% and 95\% CL,
for the $\Lambda$CDM+$\Sigma m_\nu$ model with the PCA (blue) and instantaneous (red) reionization history, derived from Planck TT+lowP.}
\label{fig:mnu}
\end{figure}

The constraint on the optical depth of the PCA reionization is $\tau=0.092\pm0.017$,
lager than that of the instantaneous reionization obtained in the 2015 Planck analysis, but with nearly the same uncertainty~\cite{Ade:2015xua}.
The reason is that the ionization fraction at high redshift makes a larger contribution to the optical depth than at low redshift.
The optical depth between any two redshifts $z_1$ and $z_2$ is an integration of $x_e$,
\begin{equation}
\tau(z_1,z_2)=0.0691(1-Y_p)\Omega_bh\int_{z_1}^{z_2} dz\frac{(1+z)^2}{H(z)/H_0}x_e(z) \,,
\label{eq:tau}
\end{equation}
where $Y_p$ is the helium abundance.
Since the Hubble parameter $H$ falls off as $(1+z)^{3/2}$ in the matter domination epoch,
the same $x_e$ can contribute more to the optical depth at higher redshift.
It is known that the amplitude of primordial spectrum of scalar perturbations $A_s$
degenerates with optical depth $\tau$ in the form $A_se^{-2\tau}$ on small scale~\cite{Ade:2013zuv},
which means that a large $\tau$ leads to a large $A_s$.
Hence a higher value of the matter fluctuation amplitude, $\sigma_8=0.838\pm0.014$,
is favored in the PCA reionization than in the instantaneous reionization,
as shown in the left panel of Figure~\ref{fig:sigma8}.
This means that such a general reionization history strengthens the tension between Planck CMB observations and
some astrophysical data, such as cluster counts and weak lensing.
As shown in the right panel of Figure~\ref{fig:sigma8},
this tension can significantly be relieved if the neutrino masses are allowed to vary,
thanks to the degeneracy between the reionization history and the neutrino masses.
Larger neutrino masses lead to a lower $\sigma_8$ through the effects of neutrino free streaming on structure formation.
The constraint on the total mass of neutrinos can be relaxed to $\Sigma m_\mu<1.03$ eV at 95\% CL in the PCA reionization scenario.
In Figure~\ref{fig:mnu} we also show the correlation of the neutrino masses with the Hubble constant.
The constraint has a broad tail to high masses, which illustrates the acoustic scale degeneracy with $H_0$.

\begin{figure}[t]
\begin{center}
\includegraphics[width=2.5in,height=2.5in]{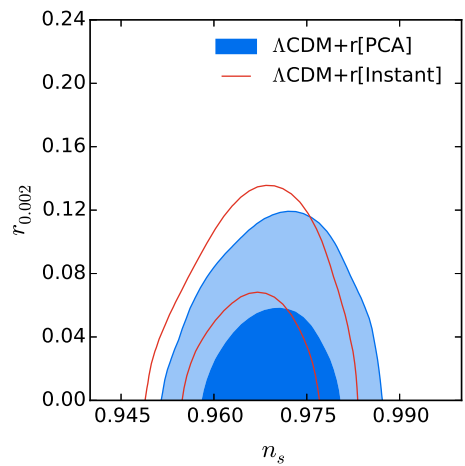}
\caption{Marginalized 2D contour in the $r_{0.002} - n_s$ plane, with 68\% and 95\% CL,
for the $\Lambda$CDM+$r$ model with the PCA (blue) and instantaneous (red) reionization history, derived from Planck TT+lowP.}
\label{fig:r}
\end{center}
\end{figure}

It is  claimed in~\cite{Pandolfi:2010dz} that the Harrison-Zel'dovich primordial spectrum (i.e., scale-invariant spectrum) of scalar perturbations
is consistent with the WMAP 7-year data in the PCA reionization scenario.
However, our analysis indicates that the Harrison-Zel'dovich spectrum is disfavored by Planck TT+lowP at 4.9 $\sigma$ CL,
although the scalar spectral index is shifted towards slightly higher values.
To test inflationary models the contribution of primordial tensor fluctuations to the CMB angular power spectra is taken into account.
Figure~\ref{fig:r} depicts the marginalized 2D contour in the $r_{0.002} - n_s$ plane, with 68\% and 95\% CL,
for the PCA (blue) and instantaneous (red) reionization history, derived from Planck TT+lowP.
We find that the allowed contour is shifted towards to higher values of $n_s$, compared to the instantaneous reionization.
This shift is important for the constraints on inflationary models.  For example,
it is  pointed out in~\cite{Ade:2015lrj} that  the class of inflationary models with a power-law potential $\phi^p$ for $p\ll 1$ and
the low-scale spontaneously broken SUSY inflationary model are marginally disfavored by Planck data at 95\% CL.
In the PCA reionization scenario, however,  these models are in agreement with recent CMB observations.
Moreover, the constraints on the tensor-to-scalar ratio is tighter compared to the instantaneous reionization
because the large amplitude of scalar spectrum suppresses the tensor-to-scalar ratio.

\section{Conclusions}

We have considered the principal component parametrization of the reionization history
and studied the effects of the PCA reionization on the estimates of cosmological parameters using recently released Planck data.
From the reconstructed reionization history we found that the Universe is not completely reionized at $z\ge 8.5$ at 95\% CL.
This conclusion can be tested by future astrophysical experiments such as SKA.
The reionization optical depth is higher than but consistent with that obtained in the instantaneous reionization scheme.
The higher value of $\tau$ leads to the higher value of $A_s$, hence the higher $\sigma_8$,
which strengthens the tension between Planck CMB observations and
some astrophysical data, such as cluster counts and weak lensing.
But the tension can significantly be relieved if the neutrino masses are allowed to vary.
The constraint on the total mass of neutrinos from CMB data is relaxed due to the degeneracy with the reionization history.
Thanks to the shift of the scalar spectral index to higher values in the PCA approach,
the low-scale spontaneously broken SUSY inflationary model,
which is marginally disfavored at 95\% confidence level in the Planck analysis, can fit the data well.

\begin{acknowledgements}
Our numerical analysis was performed on the ``Era'' of Supercomputing Center, Computer Network Information Center of Chinese Academy of Sciences.
ZKG is supported by the National Natural Science Foundation of China under Grant No.11175225 and No.11335012.
RGC is supported by the Strategic Priority Research Program of the Chinese Academy of Sciences, Grant No.XDB09000000.
We used CosmoMC and CAMB. We also acknowledge the use of Planck data.
\end{acknowledgements}

\end{document}